\documentstyle[12pt]{article}
\renewcommand
\baselinestretch{1.3}

\def\beq{\begin{equation}}
\def\brr{\begin{array}}
\def\err{\end{array}}
\def\eeq{\end{equation}}
\def\bea{\begin{eqnarray}}
\def\eea{\end{eqnarray}}
\def\bs{\bigskip}

\def\ni{\noindent}

\def\ol{\overline}
\def\nn{\nonumber}
\def\ms{\medskip}

\begin{document}

\hfill HUPD-92-10

\hfill UB-ECM-PF 92/29

\hfill October 1992

\vspace*{3mm}

\begin{center}

{\LARGE \bf
Gravitational phase transitions in infrared quantum gravity}

\vspace{4mm}

\renewcommand
\baselinestretch{0.8}
{\sc E. Elizalde}\footnote{E-mail address: eli @ ebubecm1.bitnet}
\\
{\it Department E.C.M., Faculty of Physics, University of
Barcelona, \\
Diagonal 647, 08028 Barcelona, Spain} \\  and \\
{\sc S.D. Odintsov}\footnote{On sabbatical leave from
Tomsk Pedagogical Institute, 634041 Tomsk, Russia. E-mail address:
odintsov @ theo.phys.sci.hiroshima-u.ac.jp} \\ {\it Department of
Physics, Faculty of Science, Hiroshima University, \\
Higashi-Hiroshima 724, Japan}
\ms

\renewcommand
\baselinestretch{1.4}

\vspace{5mm}

{\bf Abstract}

\end{center}

The conformal anomaly induced sector of four-dimensional quantum
gravity (infrared quantum gravity) ---which has been introduced by
Antoniadis and Mottola--- is here studied on a curved fiducial
background. The one-loop effective potential for the effective
conformal factor theory is calculated with accuracy, including terms
linear in the curvature. It is proven that a curvature induced phase
transition can actually take place. An estimation of the critical
curvature
for different choices of the parameters of the theory is given.

%\vspace{8mm}

%\noindent PACS: \begin{quote} 11.17 Theories of strings and other
%extended objects, \ 03.70 Theory of quantized fields, \
%04.50 Unified theories and other theories of gravitation.
%\end{quote}

\newpage

%\section{Introduction}

In spite of the considerable efforts invested, there has been yet no
definite success in the construction of a consistent theory of
four-dimensional (4d) quantum gravity (for a review, see [1]). A
very interesting approach for the description of 4d quantum gravity
at large distances (the so-called infrared quantum gravity) has
been developed recently by Antoniadis and Mottola [2]. Their line
of reasoning has been borrowed from that followed in two-dimensional
gravity, where the anomaly induced dynamics are already completly
different from the ones corresponding to the classical theory. In
principle, the results
of [2] provide a very natural framework for a dynamical solution of
the cosmological constant problem [3].

Let us briefly review the basic details of the effective conformal
factor dynamics (namely, the conformal sector of 4d gravity) [2]. It is
well-known that the general form of the conformal anomaly for free
conformally invariant fields is
\beq
T_{\mu}^{\mu} = b \left(C_{\mu\nu\alpha\beta}^2 + \frac{2}{3} \Box
R \right) +b'G + b'' \Box R, \ \ \ \ d=4,
\eeq
where $C_{\mu\nu\alpha\beta}$ is the Weyl tensor, $R$ the
curvature scalar and  $G$  the Gauss-Bonnet invariant. Now, as in
two dimensions [4], the conformal anomaly induces a new action.
Choosing the conformal parametrization
\beq
g_{\mu\nu} (x)= e^{2\sigma (x)} \ol{g}_{\mu\nu} (x),
\eeq
where $ \ol{g}_{\mu\nu} (x)$ is a fixed fiducial metric, one can
integrate over the conformal anomaly and get the following action
---which has been proposed as the starting point for 4d quantum
gravity [2,5]---
\bea
S_{anom}& = &\int d^4 x\, \sqrt{-\ol{g}} \left\{ b\sigma
\ol{C}_{\mu\nu\alpha\beta}^2 + 2b' \sigma \left[ \ol{\, \Box \,}^2 +
2\ol{R}^{\mu\nu} \ol{\nabla}_{\mu} \ol{\nabla}_{\nu} - \frac{2}{3}
\ol{R} \ol{\, \Box \,}+ \frac{1}{3} ( \ol{\nabla}^{\mu} \ol{R} )
\ol{\nabla}_{\mu} \right] \sigma \right. \nn \\
&+& \left. b' \left( \ol{G} -  \frac{2}{3} \ol{\, \Box \,} \ol{R} \right)
\sigma - \frac{1}{12} \left[ b'' +\frac{2}{3} (b+b') \right] \left[
\ol{R} - 6 \ol{\, \Box \,} \sigma - 6 ( \ol{\nabla}_{\mu}
\sigma )(\ol{\nabla}^{\mu} \sigma) \right]^2 \right\}.
\eea
Here $b''$ is arbitrary, since it can be changed by adding
 a local $R^2$-term to the action, and
\bea
b &=& \frac{1}{120 (4\pi)^2} (N_S+ 6N_F+12N_V-8) + \frac{199}{30
(4\pi)^2}, \nn \\
b' &=& - \frac{1}{360 (4\pi)^2} (N_S+11N_F+62N_V-28) - \frac{87}{20
(4\pi)^2}.
\eea
The contribution of the scalar ($S$), vector ($V$) and fermion
($F$) to the conformal anomaly has been known long ago (see, for
example, [8]). The last term in brackets in (4) gives the
contribution of the $\sigma$ field to the conformal anomaly [6].
The last term in both $b$ and $b'$ is the gravitation/ghost
contribution coming from the Weyl gravity action [7].

Now ---as it is done in  2d gravity--- one should add to the action (3)
the classical Einstein-Hilbert action in the parametrization (2),
thus obtaining
\beq
S_{cl} = \frac{1}{2k} \int d^4x \, \sqrt{-\ol{g}} \, e^{2\sigma}
\left( \ol{R} -6 \ol{\, \Box \,} \sigma - 6 \ol{\nabla}^{\mu}
\sigma \ol{\nabla}_{\mu}  \sigma \right) - \frac{\Lambda}{k} \int
d^4x \, \sqrt{-\ol{g}} \, e^{4\sigma}.
\eeq
Then, the full effective action is given by
\beq
S_{eff} = S_{anom}+S_{cl},
\eeq
with the additional criterion that terms independent of $\sigma$
are to be dropped out.

It has been argued in [2] ---where the study of the dynamics of the
conformal factor $\sigma$ on a flat fiducial background has been
carried out--- that this theory (which is ultraviolet
renormalizable) describes, in its infrared stable fixed point, 4d
quantum gravity at large distances (i.e., infrared quantum gravity). The
exact anomalous scaling dimension of the conformal factor at the
fixed point has been derived in [2] in the same way as is commonly done
for  2d induced gravity [9].

An indication of the necessity of an exact treatment of the
conformal factor dynamics in the infrared region comes from the
study of the graviton propagator in De Sitter
space\footnote{Recently, an interesting approach, the so-called
mean-field quantum gravity, has been discussed in De Sitter space
[12].}. It has been shown [10,11] that the graviton propagator is
not bounded at large distances, where the main contribution is
given by the conformal part [11].

Now, as  in [6], one can generalize the theory with
the action (6) to the case of a curved fiducial background, in
order to construct a multiplicatively renormalizable theory in curved
space-time:
\bea
S & = &\int d^4 x\, \sqrt{-g} \left\{  - \frac{\theta^2}{(4\pi)^2}
\sigma \Box^2 \sigma + \sigma \left[ \xi_1 R^{\mu\nu} \nabla_{\mu}
\nabla_{\nu} + \xi_2 R\Box + \xi_3 (\nabla_{\mu} R) \nabla^{\mu}
\right] \sigma  \right. \nn \\
&-& \zeta \left[ 2\alpha (\nabla_{\mu} \sigma) (\nabla^{\mu}
\sigma) \Box \sigma + \alpha^2 \left( (\nabla_{\mu} \sigma)
(\nabla^{\mu} \sigma)\right)^2 \right] + \frac{\eta_1}{\alpha^2}
e^{2\alpha\sigma} R + \eta_2 R (\nabla_{\mu} \sigma) (\nabla^{\mu}
\sigma) \nn \\
&+& \left. \gamma e^{2\alpha\sigma} (\nabla_{\mu} \sigma)
(\nabla^{\mu} \sigma) - \frac{\lambda}{\alpha^2} e^{4\alpha\sigma}
+ \frac{a_1}{\alpha^2} R_{\mu\nu}^2 + \frac{a_2}{\alpha^2} G +
\frac{a_3}{\alpha^2} R^2 + \frac{a_4}{\alpha^2} \Box R \right\}.
\eea
A few remarks are in order. First, we have dropped the bar of the
metric tensor, geometrical invariants and derivatives. Also, as in
ref. [2], we have assumed that the $e^{\sigma}$ acquires a scaling
dimension $\alpha$, and we have done the transformation $\sigma
\rightarrow \sigma \alpha$. The presence of a term linear in
$\sigma$ which cancels all terms linear in $\sigma$, both at the
classical and at the quantum level, is also assumed [2] (such a
trick is standard in quantum field theory and can be found in
textbooks [15]).

In the theory with the action (7), the coupling constants $\xi_1$,
$\xi_2$,  $\xi_3$,  $\zeta$,  $\eta_1$,  $\theta^2$, $\eta_2$,
$\gamma$, $\lambda$, $a_1$, $a_2$,  $a_3$ and $a_4$ are arbitrary.
In order to obtain a correspondence with the $S_{eff}$ given by (6),
one is bound to choose them in the following way
\bea
&& \frac{\theta^2}{(4\pi)^2} =2b+3b'', \ \ \ \zeta=2b+2b'+3b'', \
\ \ \gamma=\frac{3}{k}, \ \ \ \lambda=\frac{\Lambda}{k}, \nn \\
&& a_1=a_2=a_3=a_4=0, \ \ \ \eta_1=\frac{1}{2k}, \ \ \ \eta_2=b''+
\frac{2}{3} (b+b'), \\
&&\xi_1= 2 \left( \zeta -  \frac{\theta^2}{(4\pi)^2} \right), \ \
\ \xi_2=-\zeta+ \frac{2}{3} \frac{\theta^2}{(4\pi)^2}, \ \ \
\xi_3=- \frac{1}{3} \frac{\theta^2}{(4\pi)^2}. \nn
\eea
Notice that, given in the form (7) with arbitrary coupling
constants, this theory is multiplicatively renormalizable in curved
space-time (for a general discussion, see [1]). In the infrared
stable fixed point $\zeta =0$ [2] (the only case we shall here discuss),
 and with the parameters (8), the theory describes
qualitatively well the infrared sector of quantum gravity. Moreover, at
$\zeta =0$ the theory given by (7) is still multiplicatively
renormalizable in curved space-time.
 It is
interesting to observe also, that the theory under consideration is
probably connected with the self-dual limit of higher derivative
quantum gravity [13,14].

Of course, the model given by the action (7) contains
higher-derivative terms. Hence, it is expected to be non-unitary at
the quantum level. However, most probably the unitarity can be
restored as in the case of $c>25$ non-critical strings [16]. And,
from another point of view, only {\it fundamental} theories should
be demanded to be unitary, and not {\it effective} theories in which the
effects of spin-two gravitational modes are frozen.

We now  explicitly give the $\beta$-functions for the theory (7),
corresponding to the infrared stable fixed point $\zeta =0$. We
shall write here only the beta functions for the coupling constants
connected with the field $\sigma$ (taking into account the
classical scaling dimension for $e^{\sigma}$). The
 $\beta$-functions for the $R^2$-terms (vacuum energy)
 are not so important,
owing to the fact that the vacuum energy on a fixed background can
actually be considered as a constant. We have
\bea
\beta_{\gamma} &=& \left( 2-2\alpha + \frac{2\alpha^2}{\theta^2} \right)
\gamma, \nn \\
\beta_{\lambda} &=& \left( 4-4\alpha + \frac{8\alpha^2}{\theta^2}
\right) \lambda - \frac{8\pi^2\alpha^2\gamma^2}{\theta^4} \left( 1+
\frac{4\alpha^2}{\theta^2} + \frac{6\alpha^4}{\theta^4}  \right), \\
\beta_{\eta_1} &=& \left( 2-2\alpha + \frac{2\alpha^2}{\theta^2} \right)
\eta_1 + \frac{\alpha^2\gamma}{6\theta^2}-
\frac{(4\pi)^2\alpha^2\gamma}{\theta^4} \left(  \frac{\xi_1}{4} + \xi_2
\right), \nn
\eea
where $\beta_{\gamma}$ and $\beta_{\lambda}$ have been calculated
in [2] and  are exact results,  $\beta_{\eta_1}$ has been obtained
in the one-loop approximation [6], and the remaining
$\beta$-functions
 (i.e., those for $\xi_1$, $\xi_2$, $\xi_3$ and $\eta_2$)
do not appear in the one-loop approximation [6]. As has been argued
in ref. [2], the trace of the energy-momentum tensor for the
$\sigma$-field sector must vanish, in direct analogy with 2d
gravity [9]. Hence, the $\beta$-functions (9) must necessarily
vanish. From the conditions  $\beta_{\gamma}=0$ and
$\beta_{\lambda}=0$, one gets [2]
\beq
\alpha_{\pm} = \frac{1\pm \sqrt{1-4/\theta^2}}{2/\theta^2}, \ \ \ \
\frac{\lambda}{\gamma^2}=9k \Lambda = \frac{2\pi^2}{\theta^2} \left( 1+
\frac{4\alpha^2}{\theta^2} + \frac{6\alpha^4}{\theta^4}  \right).
\eeq
A detailed interpretation of (10) has been given in
[2]\footnote{Notice that, taking into account (10) and the values
of $\xi_1$ and $\xi_2$ from (8), equation $\beta_{\eta_1}=0$ is
fulfilled automatically}. In particular, one can understand from
this expression that the value $\theta^2_{cr}=4$ corresponds to the
$c=1$ string, where a phase transition could be expected. Thus, it
is meaningful to pose the question: is this indeed the case in the
theory under discussion? Does this model actually exhibit a phase
transition? In what follows we shall address this point.

As first step, we restrict ourselves to flat space and calculate
the Coleman-Weinberg potential for the theory (7). Let us denote
$e^{\alpha\sigma}=\phi$. Then, the one-loop effective potential for
the theory (7) is given by
\beq
V^{(1)} (\phi )= \frac{\lambda}{\alpha^2} \phi^4 + B\phi^4+A\phi^2
+ \frac{1}{2} \int \frac{d^4k}{(2\pi)^4} \left[ \ln \left( 1+
\frac{A_1\phi^2}{k^2} \right) +  \ln \left( 1+
\frac{A_2\phi^2}{k^2} \right) \right],
\eeq
where
\[
A_{1,2} = \frac{(4\pi)^2}{2\theta^2} \left( \gamma \pm \sqrt{\gamma^2-
32 \frac{\lambda \theta^2}{(4\pi)^2}} \right),
\]
and $A$ and $B$ are renormalization constants. With the same
renormalization conditions as in the paper by Coleman and Weinberg
[17], one can immediately find the values of $A$ and $B$.  Finally,
we obtain the following effective potential:
\beq
V^{(1)} (e^{\alpha \sigma}) = \frac{\lambda}{\alpha^2} e^{4\alpha
\sigma} + \frac{1}{2} \left[ \frac{\gamma^2 (4\pi)^2}{2\theta^4}-
\frac{8\lambda}{\theta^2} \right]  e^{4\alpha \sigma}  \left(
\frac{\sigma}{\sigma_0}- \frac{25}{6} \right),
\eeq
where $\sigma_0$ is the normalization point. In particular, from
(12) one can easily  obtain the value of $\sigma$ corresponding to
Coleman-Weinberg's spontaneous symmetry breaking
\beq
\frac{\sigma}{\sigma_0} = - \frac{\lambda}{\alpha^2} \left\{
\frac{1}{2} \left[ \frac{\gamma^2 (4\pi)^2}{2\theta^4}-
\frac{8\lambda}{\theta^2} \right] \right\}^{-1} + \frac{25}{6}-
\frac{1}{4\alpha},
\eeq
or, if we take into account (10),
\beq
\frac{\sigma}{\sigma_0} = - \frac{\lambda}{2(1-\alpha)} +
\frac{25}{6}- \frac{1}{4\alpha},
\eeq
where $\lambda$ is given by the second expression (10). Hence, we can
estimate
the value of $\sigma$ corresponding to spontaneous symmetry breaking.

Consider now the one-loop effective potential for the theory (7) on
a curved background. We shall only consider the linear
curvature approximation. There are different ways of doing this
calculation. One of them is to use the renormalization group method
and find the one-loop effective potential. Alternatively, we can perform
the calculations on a De Sitter background, where $\zeta$-function
regularization works very well, and expand the final answer taking
into account all terms up to the ones linear in the curvature.
Both procedures
give the same answer (provided the same renormalization conditions are
used). Since the details of such manipulations are quite standard
(for instance, they can be found in [1]), we shall here only give the
final result \bea
V^{(1)} (\phi )&=& \frac{\lambda}{\alpha^2} \phi^4 +
\frac{1}{2} \left[ \frac{\gamma^2 (4\pi)^2}{2\theta^4}-
\frac{8\lambda}{\theta^2} \right]  \phi^4 \left( \ln
\frac{\phi^2}{\mu^2} - \frac{25}{6} \right) -
\frac{\eta_1}{\alpha^2} \phi^2R \nn \\
&+& \frac{1}{2} \left[ \frac{1}{\theta^2} \left( 2\eta_1+
\frac{\gamma}{6} \right) - \frac{(4\pi)^2\gamma}{\theta^4} \left(
\frac{\xi_1}{4}+ \xi_2 \right) \right] \phi^2 R \left( \ln
\frac{\phi^2}{\mu^2} - 3 \right),
\eea
where $\phi^2\equiv e^{2\alpha\sigma}$, $\mu^2\equiv
e^{2\alpha\sigma_0}$ and we assume that $\phi^2 >>|R|$ (the linear
curvature approximation). As one can see immediately, the first two
terms represent the Coleman-Weinberg potential calculated above,
while the remaining terms provide the curvature correction.

We now concentrate on the possibility of a curvature-induced phase
transition (see [1] for a description of the general situation). An
interesting possibility is that of a first-order phase transition,
when the order parameter $<\phi >$ experiences a quick change for some
critial value of the curvature, $R_c$. The standard condition of
such a phase transition are:
\beq
V^{(1)} (\phi_c,R_c)=0, \ \ \ \left. \frac{\partial
V^{(1)}}{\partial \phi_c} \right|_{\phi_c,R_c} =0, \ \ \ \left.
\frac{\partial^2 V^{(1)}}{\partial \phi_c^2} \right|_{\phi_c,R_c}
>0.
\eeq
Let us discuss a few different cases. Impose the following (rather
natural) restrictions on
the parameters in (15): $\gamma^2 (4\pi)^2 >> |\lambda |$, $\eta_1
\simeq \gamma$, $\alpha^2 = \theta^2=1$, $\gamma <<1$, $4\pi^2
|\xi_1/4+\xi_2 |>1$ and  $4\pi^2 |\gamma (\xi_1/4+\xi_2) |<<1$.
Then it is immediate that a phase transition is possible, and that
\beq
e^{2\alpha (\sigma_c-\sigma_0)} \sim e^3, \ \ \ \ \ |R_c| e^{-
2\alpha \sigma_0} \sim - \frac{4\pi^2\gamma^2}{\eta_1 \, \mbox{sgn} \,
(R_c)} \, e^3.
\eeq
We can also consider the set of parameters (15) which correspond to
infrared quantum gravity:
\[
\frac{\gamma}{6\theta^2}- \frac{(4\pi)^2\gamma}{\theta^4} \left(
\frac{\xi_1}{4}+ \xi_2 \right) =0, \ \ \ \ \eta_1=
\frac{\gamma}{6}. \]
In particular, with the following choice: $\alpha=1$, $\theta^2 >>1$,
$\gamma <<1$, $\gamma^2/\theta^2 \sim \lambda <<1$, $\lambda > \gamma^2
\pi^2/\theta^2$, we get
\beq
e^{2\alpha (\sigma_c-\sigma_0)} \sim e^{-\theta^2/4}, \ \ \ \ \ |R_c|
e^{-2\alpha \sigma_0} \sim - \frac{32 \lambda}{\theta^2\gamma \,
\mbox{sgn} \, (R_c)} \, e^{-\theta^2/4}.
\eeq
The phase transition is clearly seen for standard values of the
parameters which satisfy these conditions (Fig. 1). It takes place for a
very reasonable value of the curvature ($R_c \simeq 1$). To be noticed
is the fact that,
also for the
critical case, $\theta^2_{cr}=4$, $\alpha_{cr}^2=2$, we do  find a phase
transition in our approach, but this happens only for sufficiently high
values of the
curvature ($R_c\simeq 100$, see Fig. 2). However, our approximation is
presumably not
good enough for very large curvature, and higher order corrections on
the curvature should be taken into account in that case.
 Also interesting is the following situation,
corresponding to the conditions in our present universe:
$\theta^2=1$ and $\alpha^2=1$. The phase transition occurs here for even
higher values of the curvature ($R_c \simeq 500$).
 It is represented in  Fig. 3.

The main qualitative result in this paper is the explicit
proof that for some values of the parameters the conformal sector
of quantum gravity can yield a curvature induced phase transition.
Also important is the observation that in the case of an external
background with torsion, the conformal sector of quantum gravity
with torsion [18] can lead to a torsion-induced phase transition.

A final remark about possible phase transitions at non-zero
temperature. It is known that the trace anomaly does not change at
non-zero temperature. Therefore, if we consider the flat fiducial
background with a compactified time dimension, the effective
potential is given again by (11), where $dk_0$ is to be changed in
the standard way. If we add to the classical potential a mass-like
term of the kind $-[m^2/(2\alpha^2)]\, e^{2\alpha \sigma}$, then $A_1$
and $A_2$ are slightly modified and we can find the critical
inverse temperature as in ref. [19]:
\beq
\beta_c^2 =- \frac{\gamma (4\pi)^2\alpha^2 }{12\theta^2m^2},
\eeq
where $m^2$ is an unphysical, negative mass [19]. Therefore, by
adding a mass-like term to the effective potential we get the
possibility of a phase transition at non-zero temperature.

In summary, we have shown that the trace-anomaly-induced dynamics
(infrared quantum gravity) has a very rich phase structure and can
undergo a phase transition induced by the curvature. It would be
interesting to understand if such a phase transition has really
taken place in the early universe and, if so, if it can lead to
some important cosmological consequences.
\vspace{5mm}

\ni{\large \bf Acknowledgments}

S.D.O. wishes to thank I. Antoniadis for extremely valuable discussions
 and the Particle Group at Hiroshima University
for kind hospitality.
S.D.O. has been supported by JSPS (Japan) and
E.E.  by DGICYT (Spain), research project
PB90-0022.
\bs

%\appendix

%\section{Appendix}

\newpage
\renewcommand
\baselinestretch{1.1}

\newpage

\renewcommand
\baselinestretch{1.3}

\ni{\large \bf Figure captions}.
\bs

\ni{\bf Fig. 1}. The effective potential at one-loop, $V^{(1)}(\phi )$
(eq. (15)), represented as a function of
 $\phi$ for several values of the curvature $R$. We have taken a typical
set of parameters corresponding to infrared quantum gravity (eq. (18)),
in particular: $\alpha =1$, $\theta^2 = 10^2$, $\gamma = 10^{-2}$ and
$\lambda = 10^{-3}$. The phase transition appears already for quite
small values of $R$ (in this example  $R_c\simeq 1$).
\ms

\ni{\bf Fig. 2}. The same as in Fig. 1, but for the  values of
constants corresponding now to the critical case:
$\theta^2_{cr}=4$, $\alpha_{cr}^2=2$, and $\lambda/\gamma^2$ as given
through
eq. (10).
The phase transition appears only for very high values of the
curvature ($R_c \simeq 100$).
 \ms

\ni{\bf Fig. 3}. The same as in Fig. 1, but for the following values of
the parameters (which correspond to our present universe):
$\theta^2=1$,  $\alpha^2=1$ and $\gamma=3/k$, where $1/2k=1/16\pi G$,
being $G$  the gravitational constant.
The phase transition appears  at even higher values of the
curvature ($R_c \simeq 500$).
 \ms


{\small

\begin{thebibliography}{99}

\bibitem{} I.L. Buchbinder, S.D. Odintsov and I.L. Shapiro, {\sl
Effective Action in Quantum Gravity}, IOP Publishing, Bristol and
Philadelphia, 1992.

\bibitem{} I. Antoniadis and E. Mottola, {\sl  Phys. Rev.} {\bf
D45} (1992) 2013.

\bibitem{} S. Weinberg, {\sl Rev. Mod. Phys.} {\bf 61}
(1989) 1.

\bibitem{} A.M. Polyakov,   {\sl Phys.
Lett.} {\bf B103} (1981) 207.

\bibitem{}  S.D. Odintsov and I.L. Shapiro, {\sl Class. Quant.
Grav.} {\bf 8} (1991) L57.

\bibitem{}  S.D. Odintsov, {\sl Z. Phys.} {\bf C54} (1992) 531.

\bibitem{} I. Antoniadis, P.O. Mazur and E. Mottola, preprint
LA-UR-92-1483, 1992.

\bibitem{} M.J. Duff, {\sl Nucl. Phys.} {\bf B125} (1977)
334; S. Deser, M.J. Duff and C.J. Isham, {\sl Nucl. Phys.} {\bf
B111} (1976) 45.

\bibitem{} V.G. Knizhnik, A.M. Polyakov and A.B. Zamolodchikov,
{\sl Mod. Phys. Lett.} {\bf A3} (1988) 819; J. Distler and H.
Kawai, {\sl Nucl. Phys.} {\bf B321}
(1989) 509;  F. David, {\sl Mod. Phys. Lett.} {\bf A3} (1988)
1651.

\bibitem{}  I. Antoniadis, J. Iliopoulos and T.N. Tomaras, {\sl
Phys. Rev. Lett.}
{\bf 56} (1986) 1319; E.G. Floratos, J. Iliopoulos and T.N.
Tomaras, {\sl  Phys. Lett.}
{\bf B197} (1987) 373; T.R. Taylor and G. Veneziano, {\sl  Phys.
Lett.}
{\bf B212} (1988) 147.

\bibitem{} I. Antoniadis and E. Mottola, {\sl  J. Math. Phys.} {\bf
32} (1991) 1037.

\bibitem{} R. Floreanini and R. Percacci, {\sl Phys. Rev.} {\bf
D46} (1992)
1566.

\bibitem{} C. Schmidhuber, preprint CALT-68-1745, 1992.

\bibitem{} E. Elizalde and S.D. Odintsov, preprint HEP-TH 9205048,
to appear in Int. J. Mod. Phys. D.

\bibitem{} T. Muta, {\sl Foundations of Quantum Chromodynamics},
World Scientific, Singapore, 1987.

\bibitem{} I. Antoniadis, C. Bachas, J. Ellis and D.V. Nanopoulos,
{\sl Nucl. Phys.} {\bf B328} (1989) 117; R.C. Myers, {\sl Phys.
Lett.} {\bf B199} (1987) 371.

\bibitem{} S. Coleman  and E. Weinberg, {\sl Phys. Rev.} {\bf D7}
(1973) 1888.

\bibitem{} I. Antoniadis and S.D. Odintsov, paper in preparation.

\bibitem{} L. Dolan and R. Jackiw, {\sl Phys. Rev.} {\bf D9} (1974)
2491.

\end{thebibliography}
}
\end{document}